# Object DUO 2: A New Binary Lens Candidate?

C. Alard[1,3], S. Mao[2], J. Guibert[1,3]

[1] DEMIRM, Observatoire de Paris, 61 Avenue de l'observatoire, F-75014 Paris, France
[2] MS 51, Center for Astrophysics, 60 Garden Street, Cambridge, MA 02138, USA
[3] Centre d'Analyse des Images de l'INSU, Batiment Perrault, Observatoire de Paris, 61 Avenue de l'Observatoire, F-75014, Paris, France

**Abstract.** We present the light curve of an unusual variable object, DUO 2, detected during the search for microlensing events by the DUO project. The star remained stable for more than 150 days before it brightened by more than two magnitudes in 6 days in the B and R bands. The light curves are achromatic during the variability. We consider possible explanations of the photometric behavior, with particular emphasis on the binary lens interpretation of the event. The masses of the lenses are quite small, with the companion possibly in the range of a brown dwarf or even a few times of Jupiter. We report evidence of blending of the source by a companion through the first detection of shift in the light centroid among all the microlensing experiments. This shift sets a lower limit of $0.3''$ on the separation between the stars. The best lens model obtained requires moderate blending, which was what motivated us to check the centroid shift that was subsequently found. The best lens model predicts a separation of $1''$ between the two blended stars. This prediction was recently tested using two CCD images taken under good seeing conditions. Both images show two components. Their separation and position angle are in good agreement with our model.

## 1. INTRODUCTION

The search for dark matter using gravitational microlensing (Paczyński 1986) has been impressively successful, with several groups discovering many microlensing candidates (Bennett et al. 1994, Alcock et al. 1993; Aubourg et al. 1993; Udalski et al. 1994a; Alard et al. 1995). One spectacular case is the first binary lensing candidate discovered by Udalski et al. (1994b), which was later confirmed by Alcock et al. (1995). Lensing by binary systems is naturally expected in about 10 percent of the lensing events (Mao & Paczyński 1991).

The DUO (Disk Unseen Objects) collaboration is a project with the main goal of searching for galactic dark matter (Alard et al. 1995). One of the microlensing candidates discovered presents a very unusual variability. In this paper, we present a detailed analysis of this event, with particular emphasis on the binary lensing interpretation, much in the same spirit of Udalski et al. (1994b).

## 2. OBSERVATIONS

### 2.1. data collection

Data presented here were collected during the 1994 observing season with the ESO 1m Schmidt telescope.

The Schmidt telescope provides photographic plates of 28cm×28cm. The scale is $67''/mm$, giving a wide field of 5.2x5.2 degrees. Two B plates (IIIaJ) and one R plate (IIIaF) were taken for each photometric night. The mean time interval between two B plates was 1.5 hours, with the R plate taken just in between.

### 2.2. reduction

All the images were reduced with the standard package of the DUO project (Alard 1995). The analysis of light curves revealed an unusual microlensing candidate. This event displays a multi-peaked structure while maintaining the achromaticity. Table 1 contains the coordinates, magnitudes, and color for this event. Because the star is close to one corner of the plate, we emphasize that the magnitude zero point could have been affected. We also caution that the colors are seriously affected by the high and variable extinction in the region surrounding the star. The 60 ×60 arcsecs region centered on the star is shown in Figure 1 (Plate ESO11307B). This can serve as a finding chart.

**Table 1.** Positions and magnitudes for DUO 2

| $RA_{2000}$ | $DEC_{2000}$ | B | R | B-R |
|---|---|---|---|---|
| 18 10 17.15 | -27 28 48.5 | 20.21 | 18.55 | 1.64 |

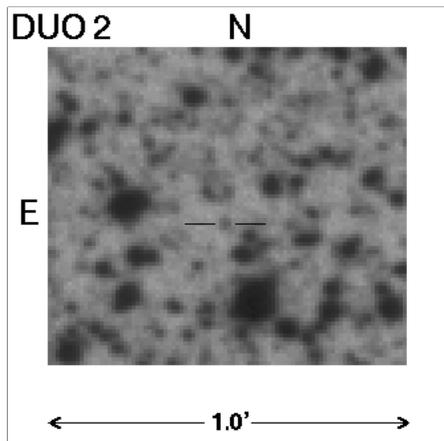

**Fig. 1.** 60 arcsec by 60 arcsec $B$-band image centered on the variable star. Two black bars identify the lensed star. North is up and east is to the left.

## 3. Nature of Variability

We now explore the possible explanations for the multi-peaked variability exhibited by this object, including variable stars and lensing by binaries.

### 3.1. An Eruptive or Flare Star?

The rapid rises and falls in the light curve suggest some type of eruptive or flare stars (cf. Udalski et al. 1994b for a similar discussion). The flare star possibility seems very unlikely as all the known flare last no more than a few hours and there is no other flaring activity in our data which may be expected for a flare star (cf. Udalski et al. 1994b). A dwarf nova or a low-mass X-ray binary (LMXB) can also explain the observed amplitude of observation. However, we note that the candidate is close to the bulge turnoff point where most stars are located, i.e., there is nothing unusual in its B-R color. While the cataclysmic variables usually have some UV excess, the U-B color for this star does not differ from the colors of the majority of the stars. Furthermore, the flat baselines in B and R are unusual for a cataclysmic variable. We also searched for the outbursts of LMXBs in the IAU circulars for the observational period, none was found in the direction of the source.

Future optical spectral observations will help to reveal the nature of the object, as dwarf novae usually have strong emission lines in the minimum state. Furthermore, if the variability repeats itself in the future, the eruptive nature of the object will be established.

### 3.2. A Double Lens?

The first binary event discovered in the direction of the bulge (Udalski et al. 1994b; Alcock et al. 1995) shows that binary lensing events can indeed be identified in the microlensing experiments, as predicted by Mao & Paczyński (1991). In comparison with the first event, our light curve has one more distinctive peak. Our "daily" time coverage translates to a worse coverage relative to the duration of variability. Nevertheless, the achromaticity in the B and R bands and the unusual variability both suggest that this may be due to a binary lens, a possibility we shall explore below.

For our binary lens model, there are 11 parameters: the total mass of the binary, $M$; the mass ratio of the individual masses, $q = m_2/m_1$; their separation, $a$, as projected onto the lens plane; the closest approach to the center of mass, $b$; the time of closest approach, $t_b$; the angle between the axis of the binary and the trajectory, $\theta$; the magnitudes of the star at the minimum light in the B and R bands, $m_B$ and $m_R$, and finally the finite size of the lensed source, $r_s$. Two more parameters, $f_R$ and $f_B$, specify the light contributions at the baselines of the lensed source in R and B. We assume the source surface brightness can be described by a limb darkening profile, $I_\lambda(r) \propto 1 - u_\lambda + u_\lambda \sqrt{1 - r^2/r_s^2}$, where $r_s$ is the radius of the star, and $u_\lambda$ is a wavelength dependent parameter (Allen 1973). We adopt $u_B = 0.6$ and $u_R = 0.5$, the values for a solar type star. The source profile has little influence on the results as it is not well constrained due to sparse sampling.

The fitting of a binary light curve has been studied by Mao & Di Stefano (1995). The fitting first finds good initial parameters and then uses these good guesses as input to search for acceptable fit(s) by minimizing $\chi^2$. For our candidate, we searched for the good initial guesses by Monte Carlo simulations. For this, we fix the baseline magnitudes at $m_R = 18.55, m_B = 20.21$, as these parameters are well determined. Furthermore, we ignore the source size. The mass ratio, separation, impact parame-ter, and angle are generated uniformly in the range of: $q = 0.1 - 1, a/r_E = 0.5 - 2.5, b/r_E = 0 - 1.25, \theta = 0 - 2\pi$. For each set of $q, a, b,$ and $\theta$, the light curve was computed. Those events with at least three peaks are further considered. For these events, we adjust the two time parameters, $t_E$ and $t_b$, such that two of the multiple peaks in the theoretical light curves coincide with the first two observed ones. The initial parameters with $\chi^2 < 348$ for 116 data points are selected for further minimization. For this final step, we vary all the parameters to search for acceptable fits.

In Fig. 2 we plot our best fit model. The model has a $\chi^2$ of 89 for 116 data points (105 degrees of freedom). The best fit parameters are:

$$q = 0.33; \ a/r_E = 1.21; \ b/r_E = 0.40; \ \theta = 94°.62;$$

$$t_E = r_E/V = 8.5 \text{ day}; \ t_b = 85.4 \text{ day}; \ f_B = 0.73; \ f_R = 0.70;$$

$$r_s/r_E \lesssim 1.0 \times 10^{-2}; m_B = 20.21; m_R = 18.55; \tag{1}$$

The model reproduces the observed light curve very well. The lensing geometry is shown in Fig. 3. The trajectory has two caustic crossings and passes close to two cusps, producing four peaks in total. The first peak is missed due to sparse sampling.

**Fig. 2.** The B and R band light curves for DUO 2. The solid line shows the light curve of the best binary microlens model. Two insets show the light curves of the lensed parts in two bands. The parameters for this fit are described in eq. (1).

### 3.3. Blending

As we showed in the last section, the best lensing model requires a moderate (30%) light contribution of an unlensed component. Originally, we only explored models without blending. For such models, we found consistent $3\sigma - 5\sigma$ deviations, especially for the point around day 85, where the magnification is $3.1 \pm 0.15$. If we assume the two adjacent peaks are produced by two caustic crossings, then such low magnification requires rather special parameters to achieve – as the absolute minimum magnification is 3 when the source is inside the caustics (Witt & Mao 1995).

**Fig. 3.** The geometry of the best fit binary microlens model. The two components are labeled as two black dots. The caustics (thick solid line) and critical curves (dashed line) are shown. The caustics are the source positions where the magnification formally diverges for a point source. The caustics on the source plane are mapped into critical curves in the lens plane. The trajectory of the source is indicated as a straight line with an arrow indicating the direction of the motion. Both axes are in units of the Einstein radius ($r_E \approx 1$ AU) in the source plane.

The blending requirement in the model prompted us to perform an additional check on the images. If the extra light is provided by a chance superimposition in the crowded field, then one may expect a positional shift in the centroid of light. The centroid position can be determined typically to 0.2 pixel (0.66″/pixel) for B and 0.35 pixel for R at the minimum state. For the bright state, the determination improves to about 0.1 pixel and 0.2 pixel, respectively. The shift of the centroid, $\Delta \boldsymbol{r_c}$, from that of the minimum state, should follow an explicit relation:

$$\Delta \boldsymbol{r_c} = (1-f)\boldsymbol{r}\,(1 - A^{-1}), \tag{2}$$

where, $\boldsymbol{r}$ is the vector from the position of the unlensed component to the position of the lensed component. In Fig. 4, we plot the positional shifts in units of pixel (0.66″) in the B band versus $(1 - A^{-1})$. By linear regression, we obtain (cf. eq. [2]).

$$(1-f)\Delta x_{c,B} = -0.41 \pm 0.09, (1-f)\Delta y_{c,B} = 0.21 \pm 0.09;$$

$$(1-f)\Delta x_{c,R} = -0.42 \pm 0.23, (1-f)\Delta y_{c,R} = 0.10 \pm 0.23, \tag{3}$$

where x increases from east to west, and y from north to south. The correlation is statistically significant in the B band but somewhat marginal in R, although the two are consistent. As $1 - f \leq 1$, eq. (3) provides a lower limit, 0.3″, on the separation between the two blended components. The orientation of the two components is 60° west of north. For the best lens model, $f \approx 0.7$, a separation of 1″ is then inferred.

After the completion of our analysis, two CCD I band images of the object were taken under seeing of 0.8″ (Van Der Hooft 1995; Szymański & Udalski 1995). The components are resolved due to the good quality of the images. The location of the unlensed star is in very good agreement with our model.

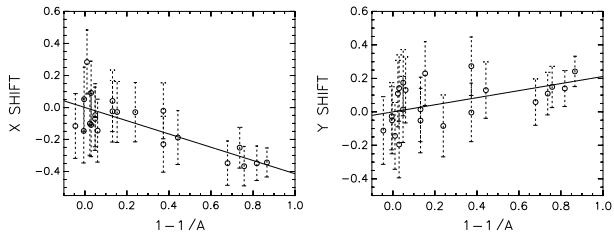

**Fig. 4.** The B band shifts in the x and y direction of the centroid of light are shown versus the $1 - A^{-1}$, where $A$ is the linear increase in flux (magnification). The shifts are expressed in units of pixel (0.66″). The solid lines are obtained by linear regression (eq. [3]).

## 4. DISCUSSION

In the previous sections we have presented the peculiar light curves of DUO 2. The binary lens model can well reproduce the observed light curve. The model requires moderate blending, which was later confirmed by the shift in the light centroid, and the CCD images. The confirmation of our prediction on the position of the blending component with the the CCD images does increase our own confidence in the lensing interpretation. The achromaticity of the light curve further supports this interpretation. Notice that although the two components are likely to have somewhat different colors, as the unlensed component is only contributing a moderate fraction of light, the achromaticity is still roughly maintained. Due to the short time duration of the variability and our short observation season, we cannot rule out the possibility that this star is just an unusual variable star, i.e., the lens model is simply a false positive (cf. Mao & Di Stefano 1995). If this event turns out to be a variable, the importance of frequent sampling is then observationally highlighted. Although we have searched a reasonable range of parameter space, we can not guarantee that the fit we presented is a unique one.

Let us now assume that the binary lens interpretation is correct and proceed to estimate the mass of the lens and the size of the source. The total mass of the lens is given by (e.g., Paczyński 1986)

$$M = 0.013 M_\odot \left(\frac{V_t}{200 \text{ km s}^{-1}}\right)^2 \left(\frac{x}{1-x}\right) \left(\frac{t_E}{8 \text{ days}}\right)^2, \quad (4)$$

where $x = D_L/D_S$ is the ratio of the distances to the lens and the lensed star, $V_t$ is the transverse velocity projected onto the source plane and we have adopted $D_S = 8$ kpc. The transverse velocity and the distance to the lens are both uncertain, but it seems likely that the lenses are low mass objects and the less massive lens can easily have the mass of a brown dwarf or even just a few times that of Jupiter. Such low mass lenses are unlikely to contribute to the observed light. The size of the lensed source is not well constrained, as the peak behaviors are not well resolved, only a rough upper limit can be obtained:

$$r_s \lesssim 2.5 R_\odot \left(\frac{V_t}{200 \text{ km s}^{-1}}\right) \left(\frac{t_E}{8 \text{ days}}\right). \quad (5)$$

Such a limit can be easily satisfied for a star like our source near the turn-off point.

Regardless of whether the variability is induced by a binary lens, we demonstrated, for the first time in all the microlensing experiments, that blending can be detected through carefully examining the shift in the centroid position.

*Acknowledgements.* It is a great pleasure to thank B. Paczyński for valuable suggestions and discussions. We also thank ESO for providing the Schmidt plates, the MAMA team for support during scanning, and E. Lesquoy for his contribution to the digitization software. We would like to thank also E. Bertin for helping the observations.